\documentclass[aps,prb,twocolumn,showpacs,amsmath,amssymb,superscriptaddress,10pt]{revtex4-1}

\usepackage[T1]{fontenc}
\usepackage[utf8]{inputenc}
\usepackage[english]{babel}

\usepackage{amssymb}
\usepackage{amsmath}
\usepackage{xcolor}

\usepackage{bbold}

\usepackage[pdftex]{graphicx}
\usepackage[colorlinks=true,citecolor=blue]{hyperref}
\usepackage{braket}

\usepackage{ulem} 
\usepackage{dsfont}
\usepackage[multidot]{grffile}

\def\up{\mathord{\uparrow}}
\def\down{\mathord{\downarrow}}

\begin{document}
\title{Iterative path-integral summations for the tunneling magnetoresistance in interacting quantum-dot spin valves}

\author{S.~Mundinar}
\author{P.~Stegmann}
\author{J.~K\"onig}
\author{S.~Weiss}
\affiliation{Theoretische Physik, Universit\"at Duisburg-Essen and CENIDE, 47048 Duisburg, Germany}

\date{\today}

\begin{abstract}
We report on the importance of resonant-tunneling processes on quantum transport through interacting quantum-dot spin valves. 
To include Coulomb interaction in the calculation of the tunneling magnetoresistance (TMR), we reformulate and generalize the recently-developed, numerically-exact method of iterative summation of path integrals (ISPI) to account for spin-dependent tunneling. 
The ISPI scheme allows us to investigate weak to intermediate Coulomb interaction in a wide range of gate and bias voltage and down to temperatures at which a perturbative treatment of tunneling severely fails. 
\end{abstract}

\maketitle

\section{Introduction}
Spintronic devices such as spin valves rely on magnetoresistive effects, in which the charge current depends on the magnetic configuration of the involved magnetic components.
Examples are the giant- (GMR) and tunnel- (TMR) magnetoresistance effect.
The TMR effect, first  observed by Julliere\cite{Julliere_1975} as a spin-dependent resistance (or conductance) in magnetic tunnel structures, is directly related to the degree of spin polarization of the tunneling electrons \cite{Meservey_1994}.
It is determined by the spin dependence of the density of states near the Fermi energy of the ferromagnetic electrodes and the tunneling matrix elements for these electrons. 
Large TMR values at room temperature of up to several hundred percent\cite{Parkin_2004,Lee_2007,Ikeda_2008} in magnetic-layer structured spin valves, also involving superconductors \cite{Vavra_2017}, have been achieved. 
For low temperatures $(T<5K)$, TMR values of about $1.9\cdot10^4\%$ have been observed in few-layer graphene heterotructures \cite{Song_2018,Wang_2018}. 
The TMR effect has already been utilized in magnetic random access memory (mRAM) devices which provide fast writing and reading as well as long lasting performance\cite{Wolf_2010}.

Quantum dots placed between ferromagnetic leads allow to realize highly tunable spin-valve devices.
Control over single spins is possible by tuning gate voltages\cite{Crisan_2015}.  
The Kondo effect and interaction-induced exchange fields have been measured\cite{Pasupathy_2004,Hauptmann_2008}, and the compensation of the exchange field by external fields has been demonstrated \cite{Hamaya_2007}. 
Controlling the magnetoresistance through carbon-nanotube quantum dots by external gates experimentally has been reported \cite{Sahoo_2005,Cottet_2006}. 
Furthermore, spin precession in non-collinear quantum-dot spin valves has been tuned by electric gates \cite{Crisan_2015}.

The theoretical description of the interplay between Coulomb interaction and spin-dependent tunneling in nonequilibrium situations imposed by a bias voltage is challenging. 
An analytical diagonalization of the underlying many-body Hamiltonian is impossible and controlled physical approximations are needed to investigate the emerging physical phenomena appropriately. Several theoretical approaches have been presented in recent years in order to describe quantum transport out of equilibrium in nanostructures.  
For weak tunnel couplings between quantum dot and leads, a perturbation expansion in the tunnel coupling is possible for various regimes of transport and gate voltages. 
A calculation to first order in the tunnel-coupling strength is sufficient to reveal the existence of an exchange field induced by finite Coulomb interaction on the quantum dot\cite{Koenig_2003,Braun_2004} or to show that the ferromagnetic Anderson-Holstein model acquires an effective attractive Coulomb interaction\cite{Weiss_2015}. 
It has been shown that the TMR through a quantum-dot spin valve to first- plus second-order in tunneling features a zero-bias anomaly \cite{Weymann_2005} that could be explained by the influence of sequential and cotunneling processes on accumulation and relaxation of the quantum spin.
A systematic scan of the TMR values (and their deviation from Julliere's value) as a function of gate and bias voltage has been performed within the same approximation scheme \cite{Weymann_2005_2}.

With increasing tunnel-coupling strengths, however, higher-order tunneling processes become more and more relevant and perturbation theory fails.
Several numerical methods have been developed to include all orders in tunneling for the Anderson model.
For instance, numerical renormalization-group approaches have been used to address the Kondo problem in the absence of a bias voltage \cite{Costi_1994}.
Picking up the idea of renormalization, functional (fRG)\cite{Khedri_2018} as well as time-dependent numerical renormalization-group theories (TD-NRG)\cite{Jovchev_2015} have been applied to the Anderson-Holstein model at finite bias voltages. 
For the interacting resonant-level model, a density-matrix renormalization group (DMRG) study has elaborated the low-energy physics and noise properties\cite{Schmitteckert_2008}. 
The time-dependent current through a dot featuring ringing effects have been studied within quantum-Monte-Carlo schemes (QMC)\cite{QMC_RMP, Muehlbacher_Komnik_2008}. 
Furthermore, the Kondo regime of the Anderson model has been approached by multilayer multiconfiguration time-dependent Hartree theory\cite{Thoss_2018}.

Some of these methods have already been used to investigate spin-dependent transport through a quantum-dot spin valve. 
It has been shown that within the linear-response regime NRG approaches work well\cite{Weymann_2011,Simon_2007,Sindel_2007}. 
Using a DMRG approach, the local density of states as well as the TMR has been calculated\cite{Gazza_2006}. 
The Kondo problem for a quantum-dot spin valve in the nonlinear response regime has been addressed by an equations-of-motion method\cite{Swirkowicz_2008}. 
In addition, the influence of the interaction-induced exchange field on dipolar and quadrupolar spin moments has been investigated \cite{Hell_2015, Misiorny_2013}. 

In this paper, we study the TMR of a quantum-dot spin valve in the resonant-tunneling regime and in the presence of both Coulomb interaction and a finite transport voltage.
For this, we adopt the numerically-exact method of iterative summation of path integrals (ISPI) \cite{Weiss_2008, Huetzen_2010, Becker_2012, Weiss_2013} and generalize it to include spin-dependent tunneling.
The ISPI scheme is formulated on the Keldysh contour and builds upon a systematic truncation of real-time correlations induced by the leads.
In this paper, we concentrate on the tunneling current in the stationary limit, although time-dependent observables and higher-order correlation functions are easily accessible within the presented approach.
The TMR depends quite sensitively on the relative importance of the various tunneling processes.
In the high-temperature regime, our results coincide with a perturbative first- or first- plus second-order calculation.
With decreasing temperature, resonant tunneling becomes more and more important.
For weak to intermediate Coulomb interaction, we reach numerical convergence of our ISPI data down to low temperatures at which perturbation theory severely fails. 
Our method, thus, provides a suitable tool to study spin-dependent, non-equilibrium quantum transport in an interacting nanostructure, covering regimes that are inaccessible by other methods.

The structure of the article is as follows.
In Sec.~\ref{sec:model} we introduce the Hamiltonian of the quantum-dot spin valve and express the Keldysh partition function within a path-integral formulation.
The Coulomb term is treated with the help of a Hubbard-Stratonovich (HS) transformation that introduces an auxiliary, Ising-like spin field.
The ISPI scheme to integrate out the HS spins is described in detail in Sec.~\ref{sec:iteration}, which also contains checks of convergence and limiting cases.
In Sec.~\ref{sec:results}, we present our results and discuss the dependence of the TMR on temperature (\ref{sec:tdepend}) as well as on bias and gate voltage (\ref{sec:biasdepend}). 
Finally, we conclude and give an outlook in Sec.~\ref{sec:conclusion}.

\section{Model \& Method}\label{sec:model}
\begin{figure}[t]
\centering
\includegraphics[width=\columnwidth]{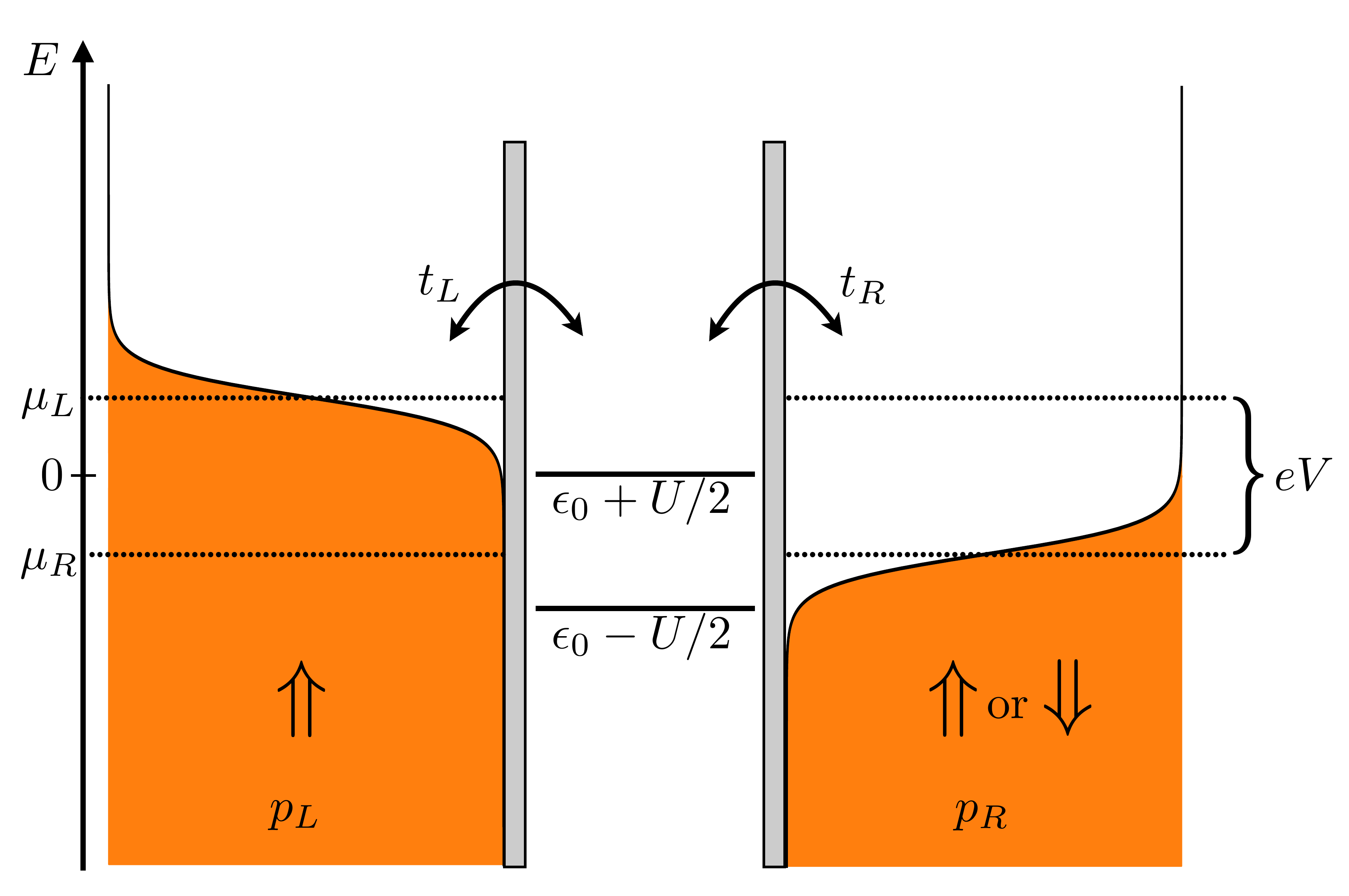}
\caption{A single level quantum dot coupled to ferromagnetic leads $\alpha=L/R$ with magnetization axes $\hat{\mathbf{n}}_L\parallel \hat{\mathbf{n}}_R$. The polarization strength $p=p_\alpha$ induces spin dependent hybridizations $\Gamma_{\alpha\tau}$. The Coulomb interaction is denoted $U$ and the single particle energy is $\epsilon_0$. A bias voltage $eV=\mu_L-\mu_R$  drops across the quantum dot. See the text for details.}
\label{fig:system}
\end{figure}

\subsection{Hamiltonian}

The Hamiltonian of a quantum dot coupled to two ferromagnetic leads (see Fig. \ref{fig:system}) consists of three parts (we set $\hbar=1$ throughout the paper), 
\begin{equation}
\mathcal H = \mathcal H_\text{dot} + \mathcal H_\text{leads} + \mathcal H_\text{T}.
\end{equation}
The first term 
\begin{equation}
\mathcal H_{\text{dot}} = \sum_{\sigma} E_0 d^\dagger_{\sigma} d_\sigma + U d^\dagger_{\uparrow} d^\dagger_{\downarrow} d_\downarrow d_\uparrow
\end{equation}
describes a single, spin-degenerate orbital of energy $E_0$, subject to Coulomb interaction $U$ when the orbital is doubly occupied.
Operators $d^\dagger_\sigma/d_\sigma$ create/annihilate an electron with spin $\sigma$ on the quantum dot, where $\sigma =\up/\down$ denotes the projection of the spin of the quantum-dot electron onto the quantization axis, which we choose along the magnetization direction of the source lead.
The energy $E_0$ is tunable via a gate voltage. 
To achieve a more compact notation, we combine the two annihilation operators $d_\sigma$ to $\mathcal{D} = (d_\uparrow, d_\downarrow)^T$ and similarly for the creation operators. 
After rewriting the interaction term by using the operator identity $2 d^\dagger_\uparrow d^\dagger_\downarrow d_\downarrow d_\uparrow = d^\dagger_\uparrow d_\uparrow + d^\dagger_\downarrow d_\downarrow -  (d^\dagger_\uparrow d_\uparrow - d^\dagger_\downarrow d_\downarrow)^2$, absorbing the quadratic part of the interaction into the single-particle energy, and defining $\epsilon_{0} = E_{0} + U/2$, we arrive at 
\begin{equation}
\mathcal H_{\text{dot}} = \mathcal{D}^\dagger \epsilon_{0}\sigma_0 \mathcal{D} - \frac{U}{2} \left(\mathcal{D}^\dagger \sigma_z \mathcal{D} \right)^2 \, .
\label{eq:Hdot}
\end{equation}
Here, $\sigma_0$ is the $2\times 2$ identity and $\sigma_z$ the $z$-Pauli matrix in spin space.

The electrons in the ferromagnetic leads $\alpha = L,R$ are described by the noninteracting Hamiltonian
\begin{equation}
\mathcal H_{\text{leads}} = \sum_{\alpha\mathbf{k}} \mathcal{C}^\dagger_{\alpha \mathbf{k}} \mathcal{E}_{\alpha \mathbf{k}} \mathcal{C}_{\alpha \mathbf{k}},
\label{eq:Hleads}
\end{equation}
with $\mathcal{E}_{\alpha \mathbf{k}} = \left(\begin{array}{cc}\epsilon_{\mathbf{k}+} -\mu_\alpha & 0 \\ 
0 & \epsilon_{\mathbf{k}-} -\mu_\alpha \end{array} \right)$, where $\epsilon_{{\mathbf k}\tau}$ is the single-particle energy for spin projection $\tau=\pm$ along the majority/minority spin direction, and $\mu_\alpha$ denotes the chemical potential.
Again, we introduced spinors $\mathcal{C}_{\alpha \mathbf{k}} = (c_{\alpha \mathbf{k}+}, c_{\alpha \mathbf{k}-})^T$ and similarly for $\mathcal{C}_{\alpha \mathbf{k}}^\dagger$.
We assume the  density of states $\rho_{\alpha\tau}=\sum_{\bf k}\delta(\omega-\epsilon_{\alpha {\bf k} \tau})$ to be constant in energy but spin dependent.
The degree of spin polarization is then characterized by $p_\alpha=(\rho_{\alpha +}-\rho_{\alpha -})/(\rho_{\alpha +}+\rho_{\alpha -})$, where $p_\alpha=0$ corresponds to a nonmagnetic lead and $p_\alpha=1$ represents a halfmetallic electrode with majority spins only.

Tunneling between quantum dot and leads is described by the Hamiltonian
\begin{align}
\mathcal H_{\text{T}} = &\sum_{\alpha \mathbf{k}} \mathcal{C}^\dagger_{\alpha \mathbf{k}} Y_\alpha \mathcal{D}   + \text{H.c.}, 
\label{eq:Htun}
\end{align}
where $t_\alpha$ is the tunnel matrix element, $Y_L=t_L \sigma_0$ as well as $Y_R=t_R\sigma_0$ for the parallel and $Y_R=t_R\sigma_x$ for the antiparallel magnetization configuration, respectively.

In this work we consider a symmetric setup, $\mu_L=-\mu_R=eV/2$, $p_L=p_R=p$, $t_L=t_R=t$, and a collinear magnetic configuration of the two leads.
Asymmetric choices for $\mu_\alpha$, $p_\alpha$, $t_\alpha$, or noncollinear magnetic configurations could straightforwardly be included in our framework.

\subsection{Keldysh functional integral}
In the following we formulate the nonequilibrium theory for the quantum-dot spin valve in terms of coherent-state functional-integrals on the Keldysh contour. The Keldysh partition functional reads
\begin{equation}
Z[\eta]=\text{Tr}\left[e^{i(S_\text{dot}+S_\text{leads}+S_\text{T}[\eta])}\right] \, .
\label{eq:ZKel}
\end{equation}
For each part of the Hamiltonian the respective action is given by
\begin{align}
S_{\text{dot}} = & \int_\mathcal{C} d t \bar{\psi}_0(t) (i \partial_t - \epsilon_0)\sigma_0 \psi_0(t)+ \frac{U}{2} \left(\bar{\psi}_0(t)\sigma_z \psi_0(t)\right)^2 \nonumber \\
S_{\text{leads}} = & \int_\mathcal{C} d t \sum_{\alpha \mathbf{k}} \bar{\psi}_{\alpha{\bf k}}(t) \left(i \partial_t \sigma_0 - \mathcal{E}_{\alpha \mathbf{k}}\right)\psi_{\alpha{\bf k}}(t) \nonumber \\
S_\text{T}[\eta] = & \int_\mathcal{C} d t \sum_{\alpha \mathbf{k}}\left[ \bar{\psi}_{\alpha \mathbf{k}} (t)Y_\alpha(t) e^{i \eta_\alpha(t)}\psi_0(t) + \text{H.c.} \right]. 
\end{align}
The path integral is calculated in the basis of fermionic coherent states, with $\psi(t),\bar{\psi}(t)$ denoting the corresponding Grassmann eigenvalue when acting with the annihilation/creation operator (spinors), as used e.g. in Hamiltonians Eqs.~\eqref{eq:Hdot} and \eqref{eq:Hleads}, onto a coherent state for lead and dot at time $t$ along the Keldysh contour $\mathcal C$, respectively\cite{Kamenev_2005, Negele_Orland}. 
We remark that the trace includes the Grassmann fields for both the quantum dot and the leads.
Since we have introduced a source term (counting field) $\eta_\alpha(t)$ to the tunneling action, we can obtain the current at time $t_m$ via the functional  derivative
\begin{equation}
I_\alpha(t_m)=\left.-i e\frac{\delta}{\delta\eta_\alpha(t_m)}\ln Z[\eta]\right|_{\eta\equiv 0}.
\label{eq:Ialpha}
\end{equation}
Note that Eq.~\eqref{eq:Ialpha} coincides with the expression for the current expectation value at measurement time $t_m$ calculated from $I_\alpha(t_m)= e \left\langle \frac{d}{dt}{\sum_{\mathbf{k}} \mathcal{C}^\dag_{\alpha{\bf k}}\mathcal{C}_{\alpha \mathbf{k}}}\right\rangle(t_m)$.

The central quantity of interest for this paper is the tunneling magnetoresistance (TMR) defined by
\begin{equation}
\text{TMR} \equiv \frac{I^\alpha_p - I^\alpha_{ap}}{I^\alpha_p} \, .
\label{eq:TMR-def}
\end{equation}
It measures the difference of the currents $I^\alpha_{(p/ap)}$ for a parallel ($p$), $\hat{{\bf n}}_L=\hat{{\bf n}}_R$, and an antiparallel ($ap$), $\hat{{\bf n}}_L=-\hat{{\bf n}}_R$, orientation of the leads' magnetization directions.
To obtain a dimensionless number between $0$ and $1$, we use as a normalization the current $I^\alpha_{p}$ for the parallel setup.
We remark that in many studies reported in the literature, the TMR is normalized with $I^\alpha_{ap}$, which nominally gives a larger value of the TMR (can be even larger than $1$) for the same physical effect.

Since the action depends quadratically on the Grassmann fields for the lead electrons, we can (without any approximation) perform the partial trace over the leads' degrees of freedom to obtain 
\begin{equation}
Z[\eta]=\text{Tr}\left[ e^{i(S_\text{dot}+S_\text{env}[\eta])}\right] \, ,
\label{eq:ZetaPsi0}
\end{equation}
The trace is now over the dot degrees of freedom only, and the environmental action
\begin{equation}
S_\text{env}[\eta]=\int_\mathcal{C}  dt \int_\mathcal{C} dt' \bar{\psi}_0 (t)\Sigma_\text{T}(t,t',\eta)\psi_0(t')
\end{equation}
involves the {\it tunneling} self energy in the presence of the counting field
$\Sigma_\text{T}(t,t',\eta)=\sum_\alpha \Sigma^{\alpha}_\text{T} (t,t',\eta)$. 
In the absence of interaction in the quantum dot, $U=0$, also the integration over $\bar{\psi}_0(t), \psi_0(t)$-fields can be performed in Eq.~\eqref{eq:ZetaPsi0}.
The resulting generating functional
\begin{equation}
Z_0[\eta]=\det\left[D^{-1}(t,t',\eta)\right] \, ,
\label{eq:ZetaU0}
\end{equation}
is expressed in terms of the Green's function of the (noninteracting) quantum dot,
\begin{equation}
	D(t,t',\eta)=\left[(i\partial_t-\epsilon_0)\sigma_0-\Sigma_\text{T}(t,t',\eta)\right]^{-1},
\end{equation}
which accounts for tunneling via the tunneling self energy $\Sigma_\text{T}(t,t',\eta)$.

We treat the Coulomb interaction term in the quantum-dot Hamiltonian by means of a Hubbard-Stratonovich (HS) transformation\cite{Hubbard_1959, Hirsch_1983} for each time $t$ along the Keldysh contour $\mathcal C$.
To enable a numerical implementation, we first discretize the Keldysh contour into $N$ time slices of length $\delta_t$ on the upper and $N$ time slices on the lower contour, respectively.
As a consequence, the Green's function $D$ becomes a $(4N\times 4N)$ matrix (the number of time slices $N$ is to be multiplied with $2$ due to spin $\sigma= \uparrow, \downarrow$ and another factor of $2$ to distinguish the upper from the lower Keldysh contour). The HS transformation relies on the identity
\begin{align}
\exp\left\{-\nu \frac{i\delta_t U}{2}\left( \bar\psi^\nu_0\sigma_z \psi^\nu_0\right)^2\right\}
=\frac{1}{2}\sum_{s=\pm1}\exp\left(-s\lambda_\nu \bar\psi^\nu_0\sigma_z \psi^\nu_0 \right)
\label{eq:HSTrafo}
\end{align}
for the contribution of the Coulomb interaction term to the action at a given time slice of length $\delta_t$ on the upper/lower Keldysh contour, $\nu= \pm$. A solution for the HS parameter $\lambda_\nu$ is (for $0 < U < \pi/\delta_t$) given by\cite{Weiss_2008, Hirsch_1983}
\begin{equation}
\lambda_\nu=\cosh^{-1}\left[\cos(\delta_t U/2)-i\nu\sin(\delta_t U/2)\right].
\end{equation}
The main idea of the HS transformation is to replace a term that is quartic in the Grassman fields by terms that are only quadratic in the Grassman fields but coupled to a newly-introduced, Ising-like degree of freedom, $s=\pm 1$.
As a result, the quantum-dot fields can be integrated out, and the resulting Keldysh generating functional 
\begin{equation}
Z[\eta]=\sum_{\{{\bf s}\}}\det\left[\Delta^{-1}[\eta]-\Sigma_C({\bf s})\right]\,,
\label{eq:zeta_1}
\end{equation}
becomes a multiple sum of determinants.
The sum runs over all $2^{2N}$ configurations of possible Ising-spin values for each time slice on the upper and lower Keldysh contour, combined in a multispin vector ${\bf s} = (s^+_1, s^-_1, s^+_2, s^-_2, \dots, s^+_N, s^-_N)$ of the HS parameters.

The matrix $\Delta[\eta]$ for the discretized Green's function of the noninteracting spin valve depends on the counting field $\eta$ but not on the HS parameters ${\bf s}$, while the matrix $\Sigma_C({\bf s})$ for the discretized self-energy due to Coulomb interaction is independent of the counting field but contains the HS parameters. 
Both are $4N\times 4N$ matrices.
To specify their matrix elements, we make use of multi-indices $a = (l,\nu,\sigma)$, where the Trotter index $l=1,\ldots, N$ labels the time slice, the Keldysh index $\nu=\pm$ distinguishes the upper from the lower Keldysh contour, and $\sigma=\pm$ labels the spin.
While the {\it charging} self energy is a diagonal matrix with matrix elements
\begin{equation}
\left( \Sigma_C({\bf s}) \right)_{aa'} = \delta_{ll'} \delta_{\nu\nu'} \delta_{\sigma \sigma'} \sigma \lambda_\nu s_l^\nu,
\end{equation} 
the discretized Green's function $\Delta[\eta]$ is diagonal in spin but nondiagonal in the Trotter and the Keldysh indices.
In the absence of the counting field, we can make use of the time translational invariance and express the matrix elements of the Green's function (making use of the multi-indices $a=(l,\nu,\sigma)$ and $a'=(l',\nu',\sigma')$) via its frequency representation,
\begin{widetext}
\begin{equation}
\left(\Delta^{-1}[0]\right)_{aa'} = \delta_{\sigma\sigma'} \int \frac{d\omega}{2\pi} e^{-i\omega (l-l')\delta_t} \left[ (\omega- \epsilon_0) \nu \delta_{\nu \nu'} 
-\frac{i}{2} \sum_\alpha \Gamma_{\alpha \sigma} \left( F_\alpha (\omega) \right)_{\nu\nu'} \right].
\label{eq:gom}
\end{equation}
\end{widetext}
Here $\Gamma_{\alpha\sigma}=2 \pi \left|t_\alpha\right|^2 \rho(\epsilon_{F\alpha\sigma})$ characterizes the hybridization of the dot with lead $\alpha$, and
$\left( F_\alpha (\omega) \right)_{\nu\nu'}$ are the matrix elements of the $2\times 2$ Keldysh matrix 
\begin{equation}
F_\alpha(\omega) =\left(\begin{array}{cc}2f_\alpha(\omega)-1&-2f_\alpha(\omega)\\ -2f_\alpha(\omega)+2&2f_\alpha(\omega)-1\end{array}\right).
\end{equation}
The Fermi function $f_\alpha(\omega)=\left[\exp(\beta(\omega-\mu_\alpha))+1\right]^{-1}$ describes the equilibrium occupation distribution of lead $\alpha$.
For later convenience, we define $\Gamma_\alpha=(\Gamma_{\alpha+}+\Gamma_{\alpha-})/2$ as the average tunnel coupling for up- and down-spin electrons.
Furthermore, we assume a symmetric coupling to the left and right lead and define $\Gamma=\Gamma_L=\Gamma_R$.

In the presence of the counting field, time translation invariance is broken and there is no simple frequency representation of $\Delta^{-1}[\eta]$ similar to Eq.~\eqref{eq:gom}.
However, since we are only interested in calculating the current (and not higher-order correlation functions), we can expand $\Delta^{-1}[\eta]$ to linear order in $\eta_\alpha$ at a single Trotter index $m$ (that corresponds to the time $t_m$ at which the current is measured) and drop the counting field anywhere else.
The resulting expression for $\Delta^{-1}[\eta]$ is given by Eq.~\eqref{eq:gom} with $\Gamma_{\alpha \sigma}$ being replaced by $\Gamma_{\alpha \sigma}[1- \frac{i \delta_t}{2}\eta_\alpha ( \nu \delta_{ml}-\nu' \delta_{ml'})]$.

Furthermore, we observe that multiplying $\Delta^{-1}[\eta]-\Sigma_C({\bf s})$ in Eq.~\eqref{eq:zeta_1} with any matrix of equal dimension that depends neither on the counting field $\eta$ nor on the HS parameters ${\bf s}$ will only result in multiplying $Z[\eta]$ with an overall factor that drops out upon calculating the current  Eq.~\eqref{eq:Ialpha}.
We make use of this freedom to increase numerical stability by multiplying $\Delta[0]$.
So, in practice, we replace the Keldysh generating functional as defined in Eq.~\eqref{eq:zeta_1} by
\begin{equation}
Z[\eta]=\sum_{\{{\bf s}\}} \det G \, ,
\label{eq:zeta_2}
\end{equation}
with 
\begin{equation}
G = \left[\Delta^{-1}[\eta]-\Sigma_C({\bf s})\right] \cdot \Delta[0] \, ,
\label{eq:G}
\end{equation}
where $\Delta^{-1}[\eta]$ has been expanded up to linear order in $\eta_\alpha(t_m)$ as explained above.

\section{Iterative summation of path integrals}
\label{sec:iteration}
\begin{figure}[t]
\centering
\includegraphics[width=\columnwidth]{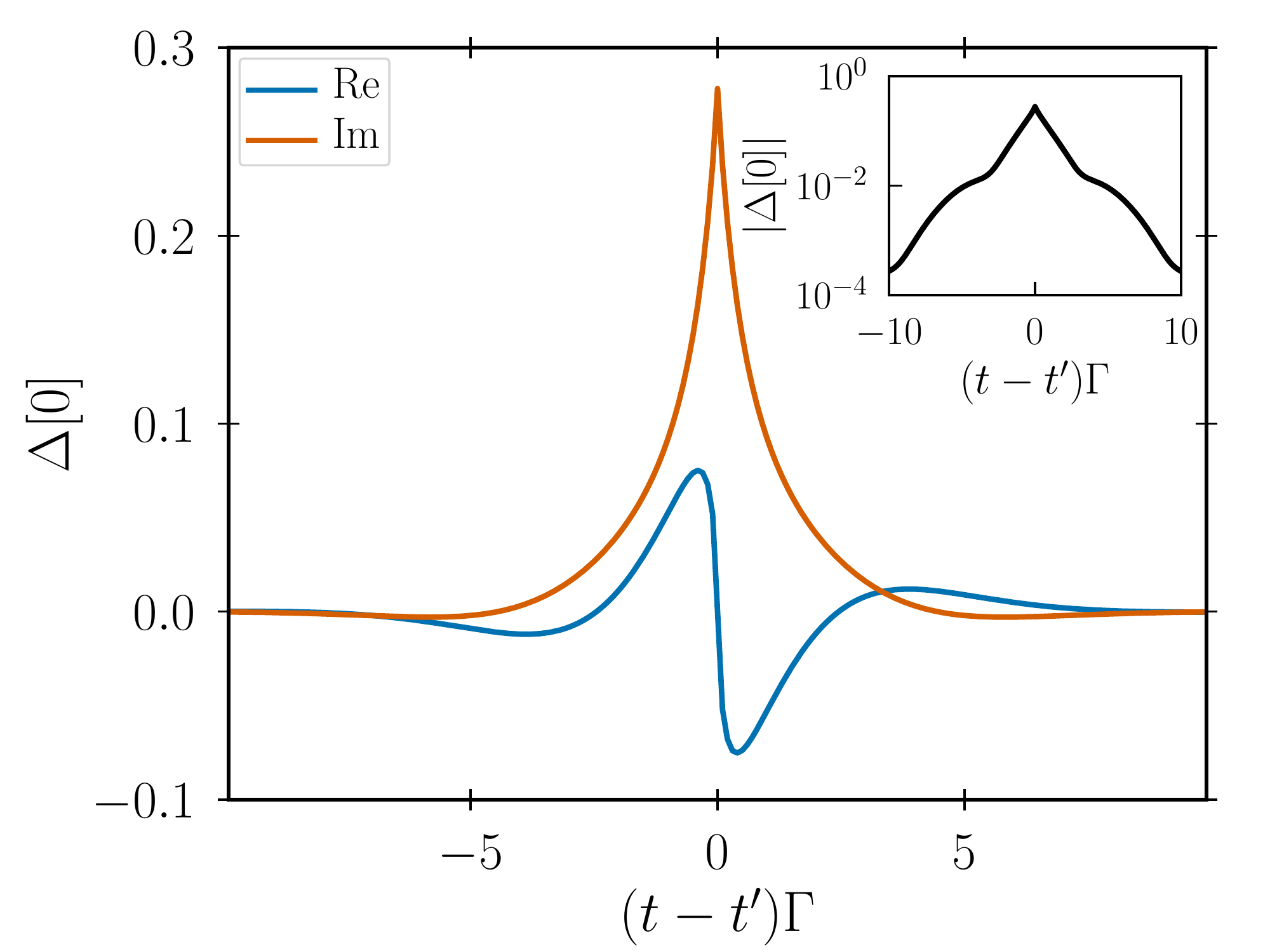}
\caption{Real and imaginary part of the matrix element of $\Delta[0]$ for $\nu=+$, $\nu'=-$, $\sigma=\sigma'=\uparrow$ as a function of $t-t'=(l-l')\delta_t$ in the continuum limit, $\delta_t \rightarrow 0$. 
Chosen system parameters are $eV=\epsilon_0=\Gamma$, $k_BT=0.2\Gamma$ and $p=0.5$ for the antiparallel setup. 
Inset: The semilogarithmic plot of the corresponding absolute value shows an exponential decay with increasing time arguments $(t-t')$.
}
\label{fig:GFtime}
\end{figure}

A brute-force summation over the $2^{2N}$ terms in Eq.~\eqref{eq:zeta_2} is a hopeless task for realistic values of $N$.
A more sophisticated and efficient way to perform the sum is needed. %, similar as it is done, e.g., in Quantum Monte Carlo simulations or DMRG calculations.
Such a method is provided by the recently-established scheme of iterative summation of path integrals (ISPI) \cite{Weiss_2008,Weiss_2013}. 
It is a completely deterministic approach, based on the iterative evaluation of traces over determinants as they appear in Eq.~\eqref{eq:zeta_2}.
The scheme relies on the fact that lead-induced correlations decrease with increasing time.
This allows us to reduce the number of summations from $2^{2N}$ to $(N/K) 2^{2K}$, where $K$ is the number of Trotter slices (of size $\delta_t$) over which correlations are taken into account.
This is a huge reduction for typical scenarios in which $N$ is of the order of a few hundreds while $K$ can be chosen to be $6$ or $7$.
We remark that a similar formulation in terms of the reduced density matrix of the system has been presented in Ref.~\onlinecite{Segal_2010}.

In Fig.~\ref{fig:GFtime}, we demonstrate how the dot's Green's function decays with time by showing the real and the imaginary part of one exemplary matrix element of $\Delta[0]$ as a function of $t-t'=(l-l')\delta_t$ in the continuum limit, $\delta_t \rightarrow 0$.
Increasing the gate and/or bias voltage, $\epsilon_0$ and $eV$, respectively, may lead to faster oscillations but does not alter the exponential decrease\cite{Weiss_2013}. 
This motivates us to neglect all contributions from time differences $t-t'$ larger than a chosen time scale $t_K$.
In the discretized version this amounts to setting all matrix elements of $\Delta[0]$ with $|l-l'| \delta_t > t_K$ to $0$. Therefore, lead-induced correlations are only included over a span of $K$ Trotter slices, where $K = t_K/\delta_t$.

The truncation after $K$ Trotter slices makes the matrix $G$ block tridiagonal.
It can, then,  be written in the form
\begin{equation}
G= \left(
\begin{array}{cccccc}
 G^{11} 	& G^{12} 	& 0 		& 0 		& \cdots & 0 \\
 G^{21}	& G^{22} 	& G^{23} 	& 0 		& \cdots & 0  \\
 0		& G^{32} 	& G^{33} 	& G^{34} 	& \cdots & 0 \\  
 0		& 0 		& G^{43} 	& G^{44}	& \cdots & 0 \\
 \vdots	& \vdots	& \vdots	& \vdots	& \ddots & \vdots  \\
 0		& 0		& 0		& 0		& \cdots & G^{N_K N_K} 
\end{array}
\right).
\label{eq:Gmatrix}
\end{equation}
The total $4N\times 4N$ matrix is decomposed into blocks $G^{nn'}$ of size $4K \times 4K$.
The index $n$ runs from $1$ to $N_K=N/K$, and only blocks $G^{nn'}$ with $|n-n'|\le 1$ have finite entries.
Since $\Sigma_C({\bf s})$ is a diagonal matrix, the definition of $G$ in Eq.~\eqref{eq:G} yields that the blocks $G^{nn'}$ in the $n$-th row depend only on the $2K$ HS spins ${\bf s}_n=(s^+_{(n-1)K+1}, s^-_{(n-1)K+1}, \ldots, s^+_{nK}, s^-_{nK})$ from the Trotter slices $(n-1)K+1$ to $nK$ but not from the ones outside this range.

To evaluate $\det G$ for a given set of HS spins, we iteratively make use of Schur's formula 
\begin{equation}
	\det \left( \begin{array}{cc} A & B \\ C & D \end{array} \right) = \det A \cdot \det ( D - C A^{-1} B ) \, .
\end{equation}
In the first step, we set $A = G^{11}$.
To calculate the determinant of $4(N-K) \times 4(N-K)$ matrix $D - C A^{-1} B$, we again use Schur's formula with the new $A$ taken as the upper left $4K \times 4K$ block.
This procedure is repeated until we arrive at the lower right corner.
The result is
\begin{equation}
Z[\eta]= \sum_{\{{\bf s}\}} \det G^{11} \cdot \det \check{G}^{22} \cdot \ldots \cdot \det \check{G}^{N_KN_K}
\end{equation}
where $\check{G}$ is iteratively defined by 
\begin{equation}
\check{G}^{nn} = G^{nn}-G^{n,n-1}\left[\check{G}^{n-1,n-1}\right]^{-1}G^{n-1,n}.
\label{eq:Gcheck}
\end{equation}
together with $\check{G}^{11} = G^{11}$.

Due to this iterative definition, $\check{G}^{nn}$ depends on all the blocks $G^{mm'}$ of Eq.~\eqref{eq:Gmatrix} that are placed above and/or to the left of $G^{nn}$, i.e. $m,m' \le n$.
This is, however, inconsistent with our decision to neglect all correlations beyond $t_K$.
Since $G$ was approximated as being block tridiagonal, we should consistently eliminate the influence of $G^{mm'}$ on $\check{G}^{nn}$ when $m,m'$ deviates from $n$ by more than $1$.
To accomplish this, we approximate $\check{G}^{nn}$ by
\begin{equation}
\tilde{G}^{nn} = G^{nn}-G^{n,n-1}\left[G^{n-1,n-1}\right]^{-1}G^{n-1,n}.
\label{eq:Gtilde}
\end{equation}
with $\tilde{G}^{11} = G^{11}$, i.e., in comparison to Eq.~\eqref{eq:Gcheck} we have replaced $\left[\check{G}^{n-1,n-1}\right]^{-1}$ by $\left[G^{n-1,n-1}\right]^{-1}$ on the right hand side.
Due to this approximation, $\tilde{G}^{nn}$ and, therefore, also its determinant $\det \tilde{G}^{nn}$ depends only on the $4K$ HS spins ${\bf s}_n$ and ${\bf s}_{n-1}$ from the Trotter slices $(n-2)K+1,\ldots, nK$ but not from earlier or later ones.
We can arrange the $2^{4K}$ different values of $\det \tilde{G}^{nn}$ for each HS spin configuration in a $2^{2K}\times 2^{2K}$ matrix, the \textit{transfer} matrix 
\begin{equation}
	\Lambda_{n,n-1} = \left( \det \tilde{G}^{nn}[{\mathbf s}_n, {\mathbf s}_{n-1}] \right)
\end{equation}
where each row corresponds to one of the $2^{2K}$ configurations for ${\mathbf s}_n$ and each column to one of the $2^{2K}$ configurations for ${\mathbf s}_{n-1}$.
The Keldysh generating functional can, then, be written as a product
\begin{equation}
Z[\eta] = \mathbf{e}^T \cdot \Lambda_{N_K,N_K-1}\cdot \ldots \cdot \Lambda_{3,2} \cdot \Lambda_{2,1} \cdot \Lambda_{1,0} 
\label{eq:iteration}
\end{equation}
of these transfer matrices, where $\Lambda_{n,n-1}$ is a $2^{2K}\times 2^{2K}$ matrix for $n\ge 2$, $\Lambda_{1,0}$ is a $2^{2K}$ dimensional column vector, and $\mathbf{e}^T = (1, \ldots, 1)$ a $2^{2K}$ dimensional row vector.
Each of the $N_K$ matrix multiplications requires a summation over $2^{2K}$ HS spin configurations. 
In total, there are $N_K 2^{2K}$ summations instead of $2^{2N}$ without the truncation scheme.
(A more precise estimate of the scaling behavior of the numerical effort should take into account the effort for building $\tilde{G}^{nn}$ from Eq.~\eqref{eq:Gtilde}.)

To illustrate the numerical effort, we mention that our code runs for given parameters and $K=6$ typically one hour on a linux cluster with $16$ nodes when the current is measured at $\Gamma t_m=10$. For finer discretizations we use a CrayXT6m machine where data points including e.g. $K=7$  are calculated within $2.5$ hours on $240$ nodes. 

\subsection{Convergence}
\begin{figure}[ht!]
\centering
\includegraphics[width=\columnwidth]{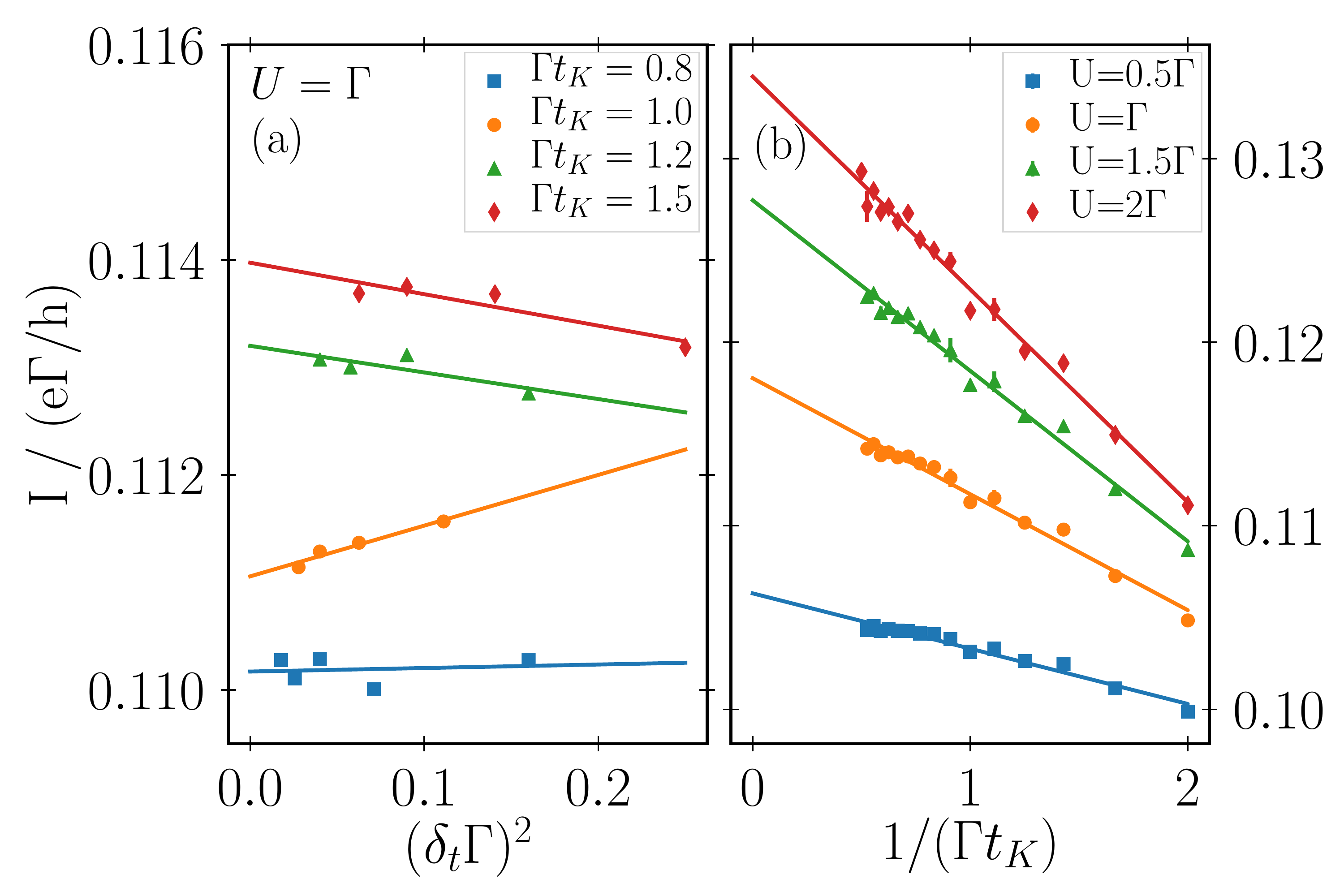}
\caption{(a) Current as a function of time step $\delta_t$ for different values of the correlation time $t_K$ and fixed interaction strength $U=\Gamma$. (b) Dependence of the current on the correlation time $t_K$ in the limit $\delta_t\to 0$ for various Coulomb interactions strengths.
Other parameters in both panels are $\hat{\mathbf{n}}_L=\hat{\mathbf{n}}_R, p=0.5, eV=\epsilon_0=\Gamma$ and $k_BT=0.2\Gamma$.}
\label{fig:convergence}
\end{figure}
The discrete HS transformation Eq.~\eqref{eq:HSTrafo} needs finite discretization time steps $\delta_t$.
This introduces a finite Trotter error for the calculation of the path integral along the Keldysh contour\cite{Weiss_2004,Hirsch_1983}, which scales with
$\delta_t^2$ for $\delta_t\to 0$.
A second source of systematic error comes from truncating correlations beyond $t_K$.
To reduce both errors and estimate the remaining error bars, we perform the following three-step procedure.

(i) First, we choose a finite correlation time $t_K$.
For this $t_K$, the time step is varied and we perform the well-established extrapolation\cite{Weiss_2004} to the continuum limit, $\delta_t\to 0$. 
A typical outcome of this procedure is shown in panel (a) of Fig.~\eqref{fig:convergence}.
We show the current for $\hat{\mathbf{n}}_L=\hat{\mathbf{n}}_R$ for different correlation times $\Gamma t_K=0.8,1.0,1.2,1.5$ and fixed Coulomb interaction $U=\Gamma$ as a function of the Trotter step size. 
In this example we have chosen the other parameters as $eV=\epsilon_0=\Gamma, p=0.5, k_BT=0.2\Gamma$. 
ISPI data are given as symbols, and the solid lines display the linear regression, which is used for the extrapolation towards $\delta_t\to 0$.
This step eliminates the Trotter error.

(ii) We repeat the extrapolation $\delta_t\to 0$ for different finite values of the correlation time $t_K$.
To eliminate the respectively associated truncation error, we extrapolate to infinite correlation time with the help of a linear regression for $1/(\Gamma t_K)\to 0$.
This is illustrated in panel (b) of Fig.~\eqref{fig:convergence} for different values of the Coulomb interaction $U$.

(iii) Finally, the error bars of the ISPI data are estimated from the standard deviations of both subsequent linear regressions. 

Since we are interested in the stationary values of the current (and the TMR), we need to set the measurement time $t_m$ large enough to overcome transient behavior.
We find that $\Gamma t_m\geq 10$ is sufficient.

For the noninteracting case, $U=0$, the errors are eliminated very efficiently with the described extrapolations, such that error bars are smaller than the symbol size in all curves presented for this case.
With increasing $U$, the error bars become larger.
We are able to obtain reliable results within reasonable computation time for intermediate values of the interaction up to $U=2 \Gamma$. 
Whenever the orbital of the dot is tuned far outside the transport window opened by the bias voltage (large $|\epsilon_0|$ or small $eV$), the ISPI data for the current and the TMR become noisy.

\subsection{Benchmark 1: noninteracting case}

To convince ourselves of the reliability of our ISPI code, we perform two benchmark tests.
First, we consider a noninteracting spin valve, $U=0$.
In this case, the current and, thus, the TMR can be calculated analytically.
The formulae for the current in the parallel and antiparallel configuration are given in Appendix \ref{appA}.

\begin{figure}[t]
\centering
\includegraphics[width=\columnwidth]{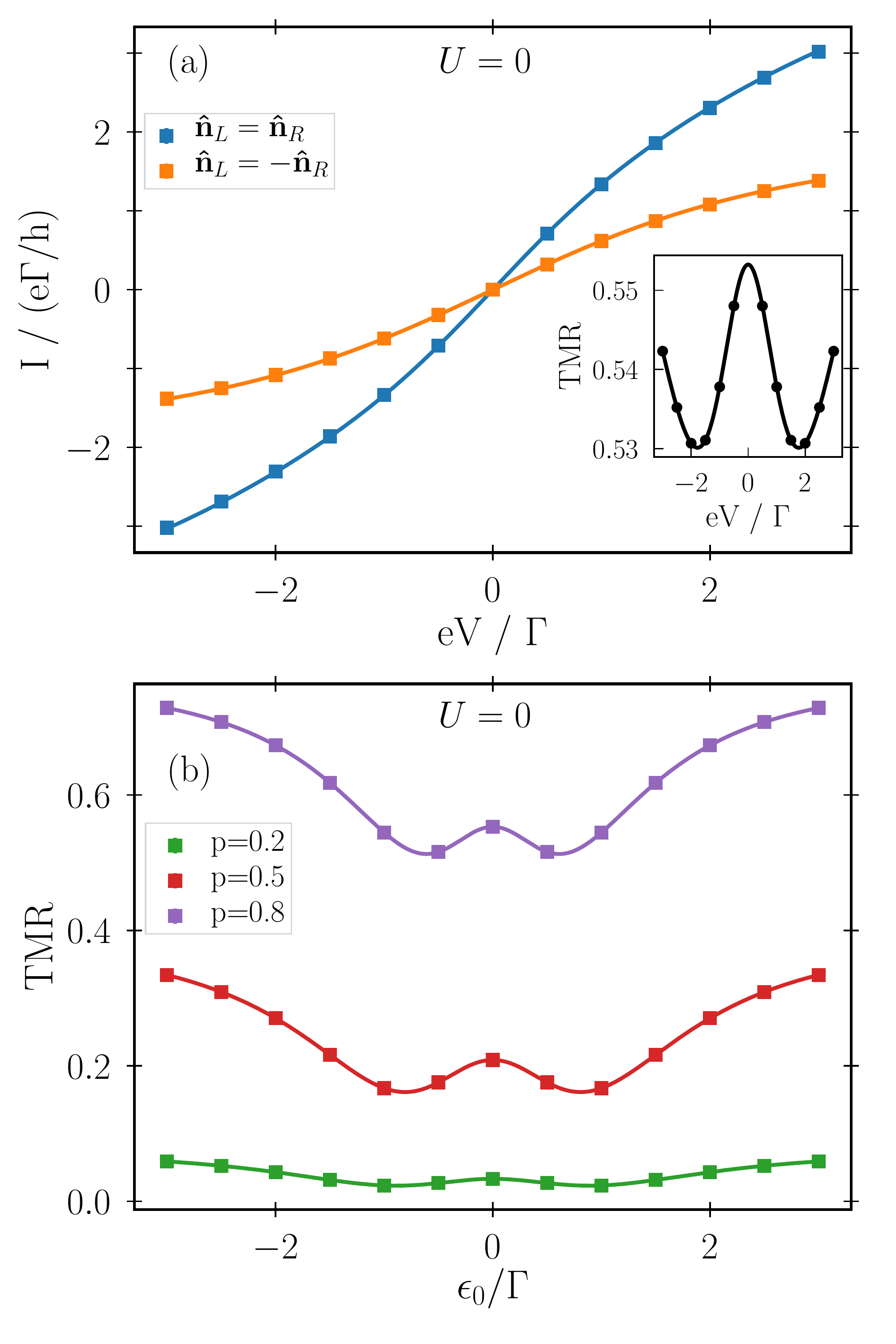}
\caption{(a) $I(V)$-characteristics for the noninteracting system for $\epsilon_0=0, p=0.8,k_BT=0.2\Gamma$ in the parallel and antiparallel configuration, respectively. The inset shows the resulting TMR. Panel (b) depicts the dependence of the TMR on gate voltage $\epsilon_0$ for different values of $p$ at $k_BT=0.2\Gamma$ and $eV=0.1\Gamma$. All error bars are of the order of the symbol size.
\label{fig:U0current}}
\end{figure}

In Fig.~\ref{fig:U0current}(a), we show the current through a noninteracting ($U=0$) quantum-dot spin valve for the parallel (blue) and the antiparallel (orange) configuration as a function of the bias voltage.
ISPI data are represented by the symbols, with error bars of the order of the symbol size and solid lines display the analytical results.
The parameters are chosen as $\epsilon_0=0, p=0.8, k_BT=0.2\Gamma$.
We find excellent agreement for the full $I(V)$ characteristics. For other parameters we have checked that the analytical curves are reproduced by the ISPI data with high precision as well.

The inset of Fig.~\ref{fig:U0current}(a) shows the TMR.
There is a well-pronounced zero-bias peak, which is well-resolved by the ISPI data.

In Fig.~\ref{fig:U0current}(b), we depict the TMR as a function of the level position $\epsilon_0$ for different lead polarizations $p$. 
Again, the ISPI data (symbols) are in perfect agreement with the analytical solution (solid  lines). 
There is a pronounced peak at $\epsilon_0=0$, again well-resolved by the ISPI data.
For increasing $p$, the current for the antiparallel lead configuration decreases, which results in a larger TMR, see Eq.\eqref{eq:TMR-def}.

\subsection{Benchmark 2: perturbation theory}
 
\begin{figure}[ht!]
\centering
\includegraphics[width=\columnwidth]{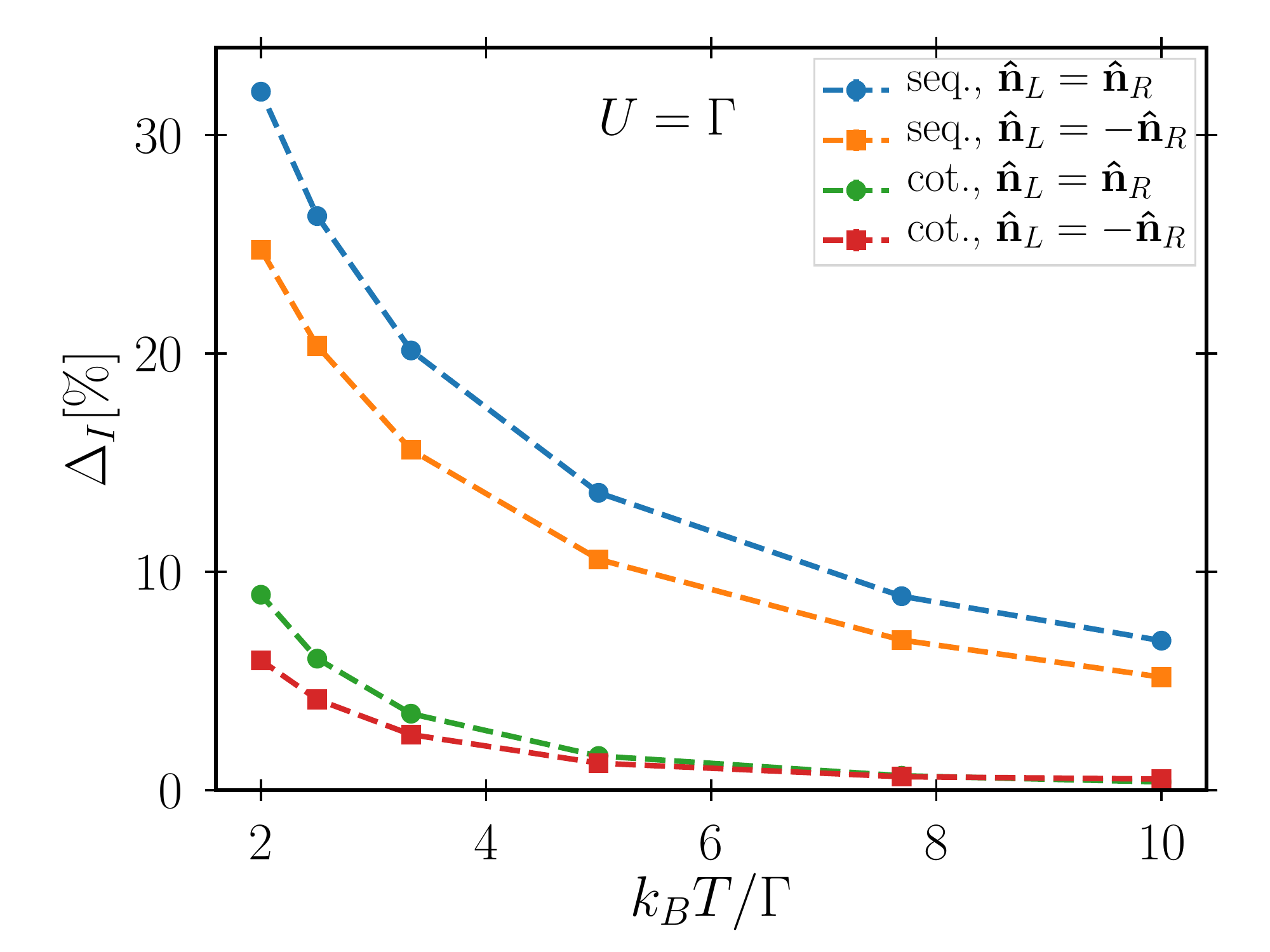}
\caption{Relative deviation of the current obtained by perturbation theory to first order (seq) or first- plus second order (cot) from ISPI as a function of temperature. Other parameters are $eV=3\Gamma, \epsilon_0=0, U=\Gamma, p=0.5$, lines are guides to the eye.
\label{fig:seqTun}}
\end{figure}

In the presence of interaction, a full analytical solution is not available anymore.
In the limit of high temperature, however, the tunnel coupling can be treated in perturbation theory.
To lowest order, only sequential-tunneling processes contribute.
The first-order theory for a quantum-dot spin valve formulated within a diagrammatic real-time technique\cite{Braun_2004} has been extended to a full first- plus second-order (referred to as sequential plus cotunneling) calculation\cite{Weymann_2005}.

This enables us to perform a second benchmark test.
We compare the ISPI data with those obtained from first- or first- plus second-order perturbation theory as a function of temperature and check whether the ISPI data converge to the perturbative results at high temperatures.
The result is shown in Fig.~\ref{fig:seqTun}, where we plot the relative deviation
\begin{equation}
	\Delta_I = \left| \frac{I^\text{seq/cot}-I^\text{ISPI}}{I^\text{ISPI}} \right|
\end{equation}
of the first-order (seq) or first- plus second-order (cot) current from the ISPI results as a function of temperature.
For the perturbative results, we use the scheme developed in Ref.~\onlinecite{Weymann_2005}.
We cover both the parallel and the antiparallel configuration.
Other parameters are set to $eV=3\Gamma, \epsilon_0=U=\Gamma, p=0.5$ and lines are guides to the eye only.

We find the ISPI data in accordance with the perturbative calculation for $k_BT\gg \Gamma$.
While a first-order description is only sufficient for temperatures much larger than $\Gamma$, the first- plus second-order calculation remains reliable for temperatures that are somewhat larger than $\Gamma$.
But for $k_BT\lesssim \Gamma$, perturbation theory clearly fails and cannot compete with ISPI.

\section{Results}
\label{sec:results}
We now turn to the discussion of the TMR through a quantum-dot spin valve for the experimentally relevant regime in which the energy scales for Coulomb interaction, tunnel coupling, and temperature are all of the same order of magnitude.
Neither the fully analytical $U=0$ formulae, nor a perturbative treatment of tunneling, nor any zero-temperature calculation done with other methods such as NRG is applicable.

The TMR through a quantum-dot spin valve depends quite sensitively on the relative importance of different tunnel processes that contribute directly and indirectly to the current.
A reference is set by Julliere's value
\begin{equation}
	\text{TMR}^\text{Jull} = \frac{2 p^2}{1+p^2}
\end{equation}
describing a direct tunnel junction between two ferromagnets without a quantum dot placed in between.
(We remind that, in our definition \eqref{eq:TMR-def} of the TMR, we normalize with the current for the parallel configuration.
The more often used normalization with the current for the antiparallel configuration would give $\text{TMR}^\text{Jull} = 2 p^2/(1-p^2)$.)
This value is easily derived from the fact that the transmission through the junction for a given spin is proportional to the corresponding density of states in the source and the drain, respectively, and the latter are proportional to $1+p$ for the majority and $1-p$ for the minority spins.
For the quantum-dot spin valve, the TMR is strongly influenced by the possibility to accumulate a finite spin polarization on the quantum dot.
The spin accumulation has the tendency to weaken the spin-valve effect, i.e., to reduce the TMR.
The value of the dot's spin polarization, on the other hand, results from an interplay of various processes that either lead to a spin accumulation or that provide a relaxation channel for the accumulated spin.
This makes the TMR a highly nontrivial function of temperature, gate- and bias voltage, and Coulomb interaction.
Since tunnel processes of different orders in the tunnel coupling contribute very differently to spin accumulation and spin relaxation, we are not surprised that a perturbative calculation of the TMR breaks down already at relatively high temperatures.

The fact that the TMR depends rather sensitively on the nature of the dominating transport channel has been extensively discussed in the literature.
Measurements of the TMR for weak tunnel coupling outside the Coulomb-blockade regime \cite{Crisan_2015} could be explained within a sequential-tunneling picture.
In the Coulomb-blockade regime, however, cotunneling dominates, leading to an enhancement of the TMR as compared to the sequential-tunneling result \cite{Weymann_2005,Weymann_2005_2,Gazza_2006,Weymann_08,Martinek_02}.
For a qualitative explanation of the TMR measured through a quantum-dot spin valve with strong tunnel coupling, a Breit-Wigner form of the transmission has been assumed \cite{Sahoo_2005,Hamaya_2007_2}.
While the Breit-Wigner form takes resonant-tunneling processes into account, it fails to describe correlations that, at very low temperature, give rise to the Kondo effect.
The formation of Kondo correlations is accompanied with an enhancement of the transmission through the dot, which gives rise to a large, bias-voltage-dependent TMR as observed in Refs.~\onlinecite{Hauptmann_2008,Pasupathy_2004,Hamaya_2007}.
Finally, we mention that also for quantum-dot spin valves in which not only the leads but also the island is ferromagnetic, an enhancement of the TMR due to cotunneling has been experimentally \cite{Ono_97, Schelp_97,Yakushiji_01} and theoretically \cite{Takahashi_98} found.

\subsection{Temperature dependence of TMR}
\label{sec:tdepend}
\begin{figure}[ht!]
\centering
\includegraphics[width=\columnwidth]{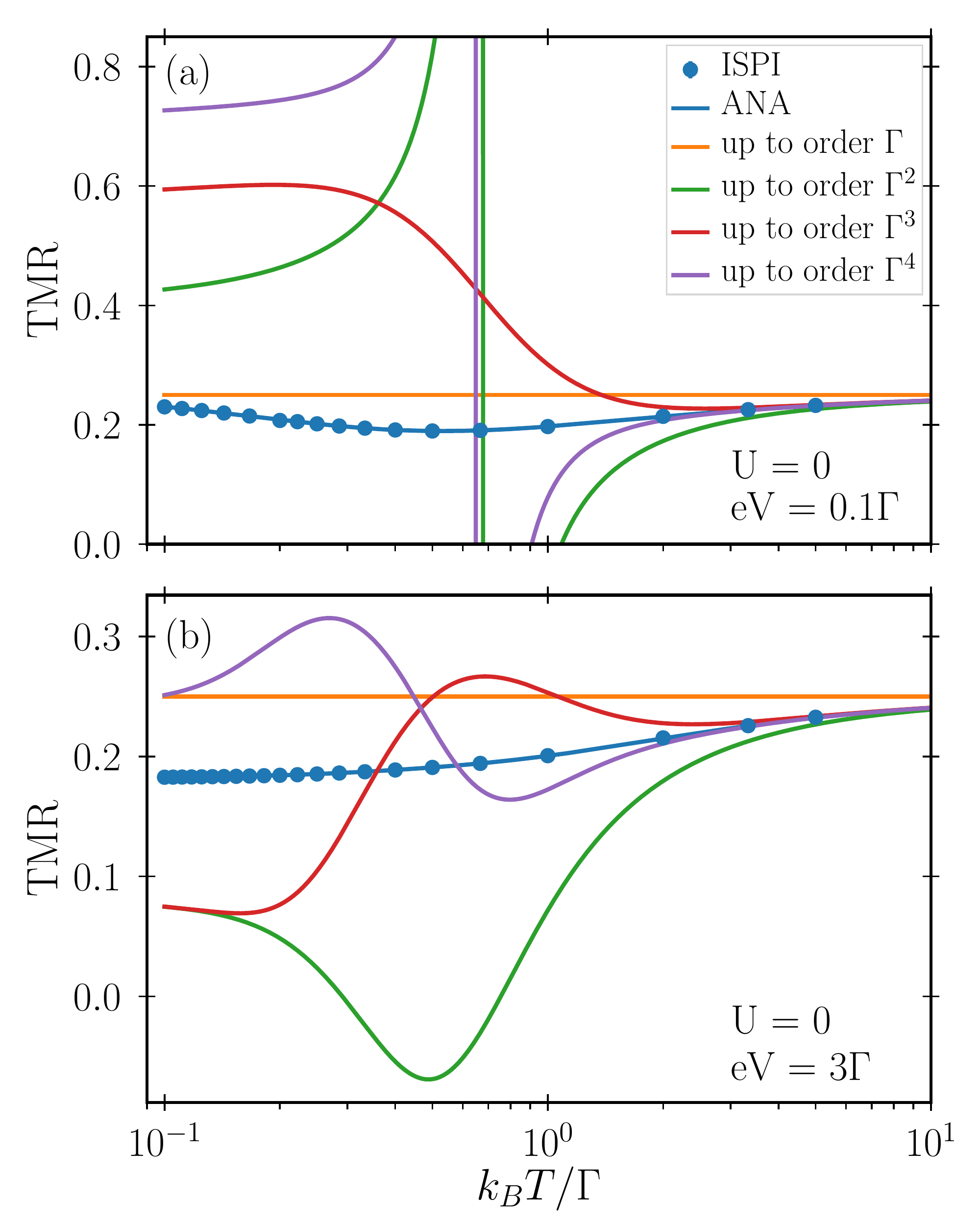}
\caption{\label{fig:res_tun_bet}Temperature dependence of the TMR for the noninteracting system, $U=0$. (a) Low-bias voltage regime,   $eV=0.1\Gamma$, (b) large-bias voltage regime, $eV=3\Gamma$. Other parameters are $\epsilon_0=0, p=0.5$. ISPI data error bars are smaller than the symbol size.}
\end{figure}

The relative importance of resonant tunneling, i.e. higher-order tunneling contributions, as compared to sequential and cotunneling is nicely demonstrated by the temperature dependence of the TMR.
In Fig.~\ref{fig:res_tun_bet}, we show the results for a noninteracting quantum-dot spin valve, $U=0$, both (a) for the linear-response, $eV=0.1 \Gamma$, and (b) the non-linear response regime, $eV= 3\Gamma$.
The ISPI data (symbols) perfectly agree with the analytical results (ANA) available for the noninteracting case.
To distinguish the contributions from different orders in tunneling, we expand the analytical result up to order $\Gamma$, $\Gamma^2$, $\Gamma^3$, and $\Gamma^4$, respectively.

To lowest order, corresponding to the sequential-tunneling approximation, the TMR is temperature independent\cite{Weymann_2005}. 
Including the second- (cotunneling), third- or fourth-order contribution makes the TMR a nonmonotonic function of the temperature.
Unphysical oscillations arise which, for small bias voltage, even extend to TMR values outside the interval $[0;1]$.
The cotunneling approximation seems to give reliable results for temperatures somewhat smaller than $k_BT\approx 10\Gamma$.
Strikingly, even a fourth-order expansion of the full analytical result is severely failing to describe the correct temperature dependence of the TMR down to $k_BT \lesssim \Gamma$.
This demonstrates the overwhelming importance of resonant tunneling when temperature and $\Gamma$ are of the same order of magnitude.
Only for $k_BT\gg 10 \Gamma$ (not shown in the figure), the ISPI data converge to the sequential-tunneling result.

\begin{figure}[ht!]
\centering
\includegraphics[width=\columnwidth]{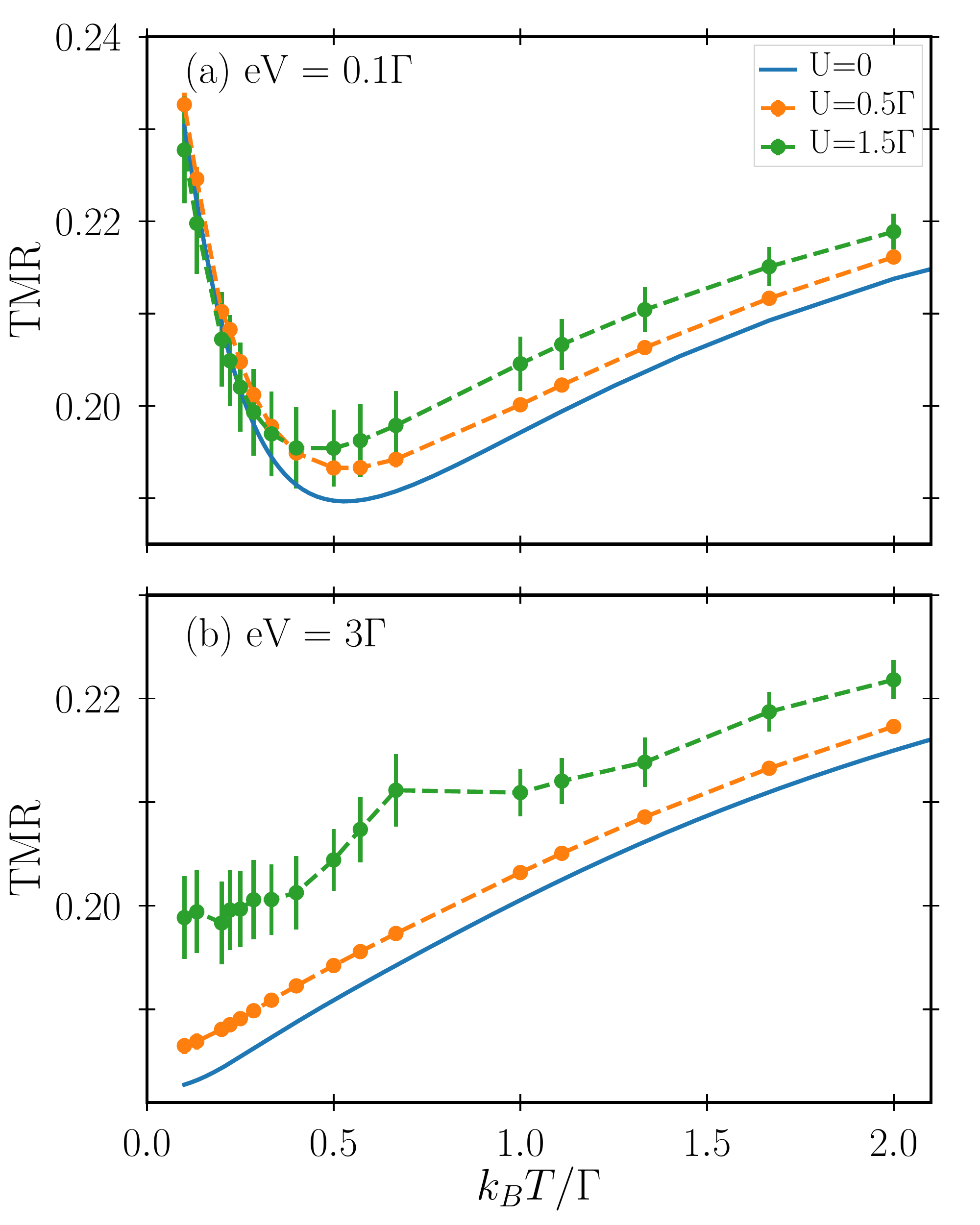}
\caption{\label{fig:TMRfinU}Temperature dependence of the TMR for $U/\Gamma=0$, $0.5$, and $1.5$. (a) Low-bias voltage regime,   $eV=0.1\Gamma$, (b) large-bias voltage regime, $eV=3\Gamma$. Other parameters are $\epsilon_0=0, p=0.5$.}
\end{figure}

We proceed with presenting the temperature dependence of the TMR for an interacting quantum-dot spin valve in Fig.~\ref{fig:TMRfinU}.
A full analytical solution to compare with is not available anymore.
We also refrain from showing results from a first- plus second-order calculation since the resulting TMR values are far off for the considered temperature regime.
Instead, we compare the TMR for different values of $U$.
We use the same parameters as for the non-interacting case and, again, show both the linear- and nonlinear-response regime.
The TMR increases with increasing Coulomb interaction $U$.

In the linear-response regime, Fig.~\ref{fig:TMRfinU}(a), we find a non-monotonic dependence of the TMR on temperature.
For $T\rightarrow \infty$, the TMR approaches the sequential-tunneling value $\text{TMR}=p^2=0.25$, as required. 
Interestingly, for $\epsilon_0=0$ and small bias voltage, also the low-temperature limit, $T\rightarrow 0$, yields $\text{TMR}=p^2$. 
In between, around $k_BT \approx 0.5 \Gamma$, the TMR displays a local minimum with TMR $\approx 0.195$. 

In Fig~\ref{fig:TMRfinU}(b), we present the temperature-dependent TMR for the nonlinear bias regime, $eV=3\Gamma$.
In this case, the TMR increases monotonically with temperature and approaches the expected sequential-tunneling value $p^2$.

\subsection{Gate- and bias-voltage dependence of TMR}
\label{sec:biasdepend}
\begin{figure}[ht!]
\centering
\includegraphics[width=\columnwidth]{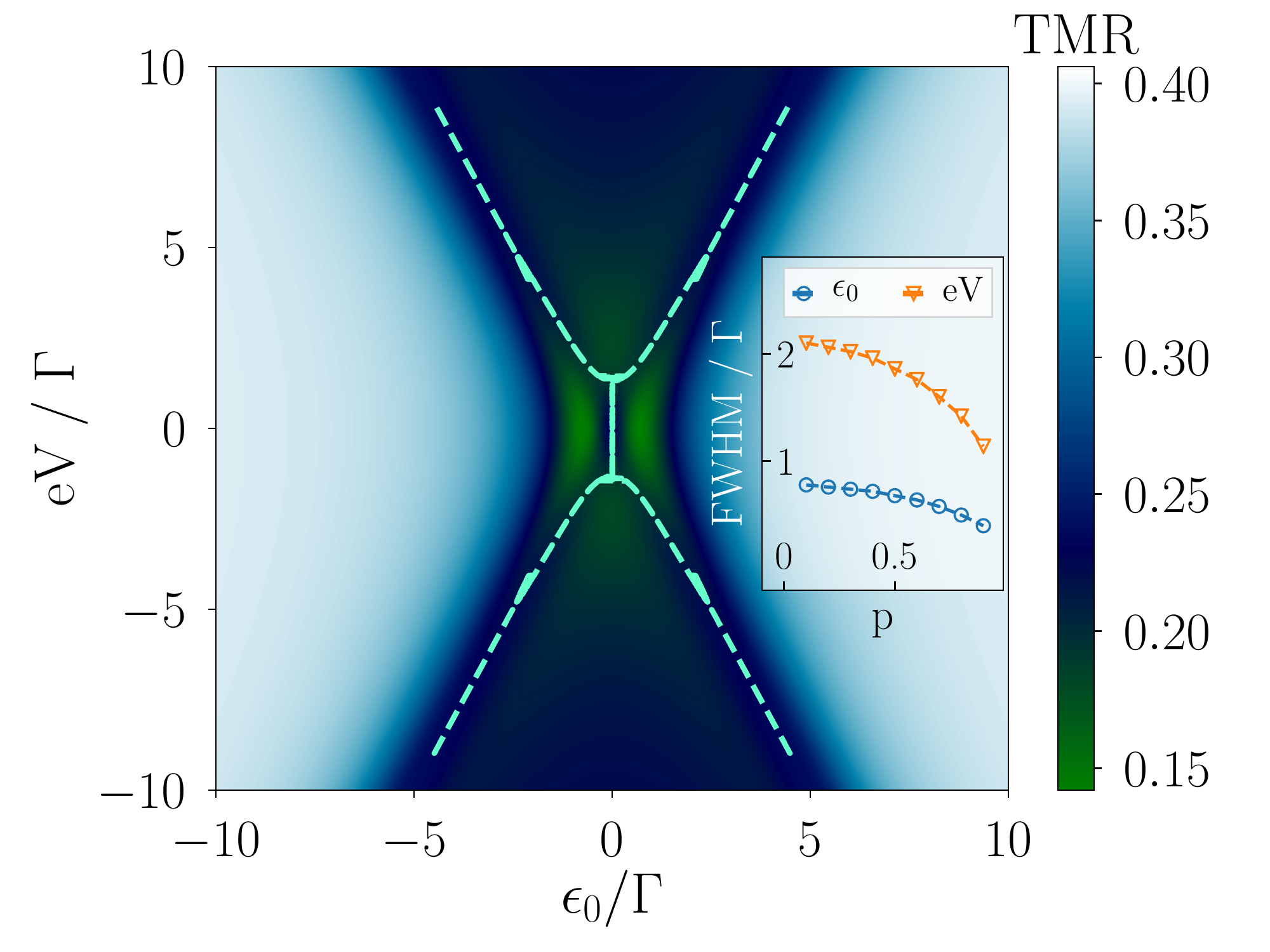}
\caption{\label{fig:TMRdens} TMR for the noninteracting system as a function of gate and bias voltage. The temperature is $k_BT = 0.1\Gamma$ and leads' polarization is given by $p=0.5$. The dashed line highlights the position of the local maximum of the TMR as a function of $\epsilon_0$. We find a bifurcation at finite bias voltage. (inset) Full width half maximum of the local maximum in the center along $\epsilon_0$- and $eV$-axis as a function of the polarization $p$.}
\end{figure}

We proceed with analyzing the gate- and bias-voltage dependence of the TMR since both gate- and bias voltages are usually tuneable in experiment with high precision.
First, we consider the noninteracting case, $U=0$.
At low temperature, $k_BT=0.1\Gamma$, the TMR shows a rather rich structure as a function of both voltages, see Fig.~\ref{fig:TMRdens}.
In particular, there is a peak at $\epsilon_0=0$, $V=0$.
The full width at half maximum of the peak as a function of $\epsilon_0$ and $V$, respectively, depends on the degree of spin polarization $p$, see inset of Fig.~\ref{fig:TMRdens}.
With increasing bias voltage, the local maximum of the TMR as a function of $\epsilon_0$ first stays at $\epsilon_0=0$ but then splits into two local maxima.
In Fig.~\ref{fig:TMRdens}, the position of the local maximum as a function of $\epsilon_0$ is highlighted by a dashed line.
Once the level position $\epsilon_0$ is tuned far away from resonance, $|\epsilon_0|\gg \Gamma$, the TMR approaches Julliere's value $2p^2/(1+p^2)=0.4$.
This is expected since in this regime, accumulation and relaxation of the quantum-dot spin can be neglected. 

\begin{figure}[ht!]
\centering
\includegraphics[width=\columnwidth]{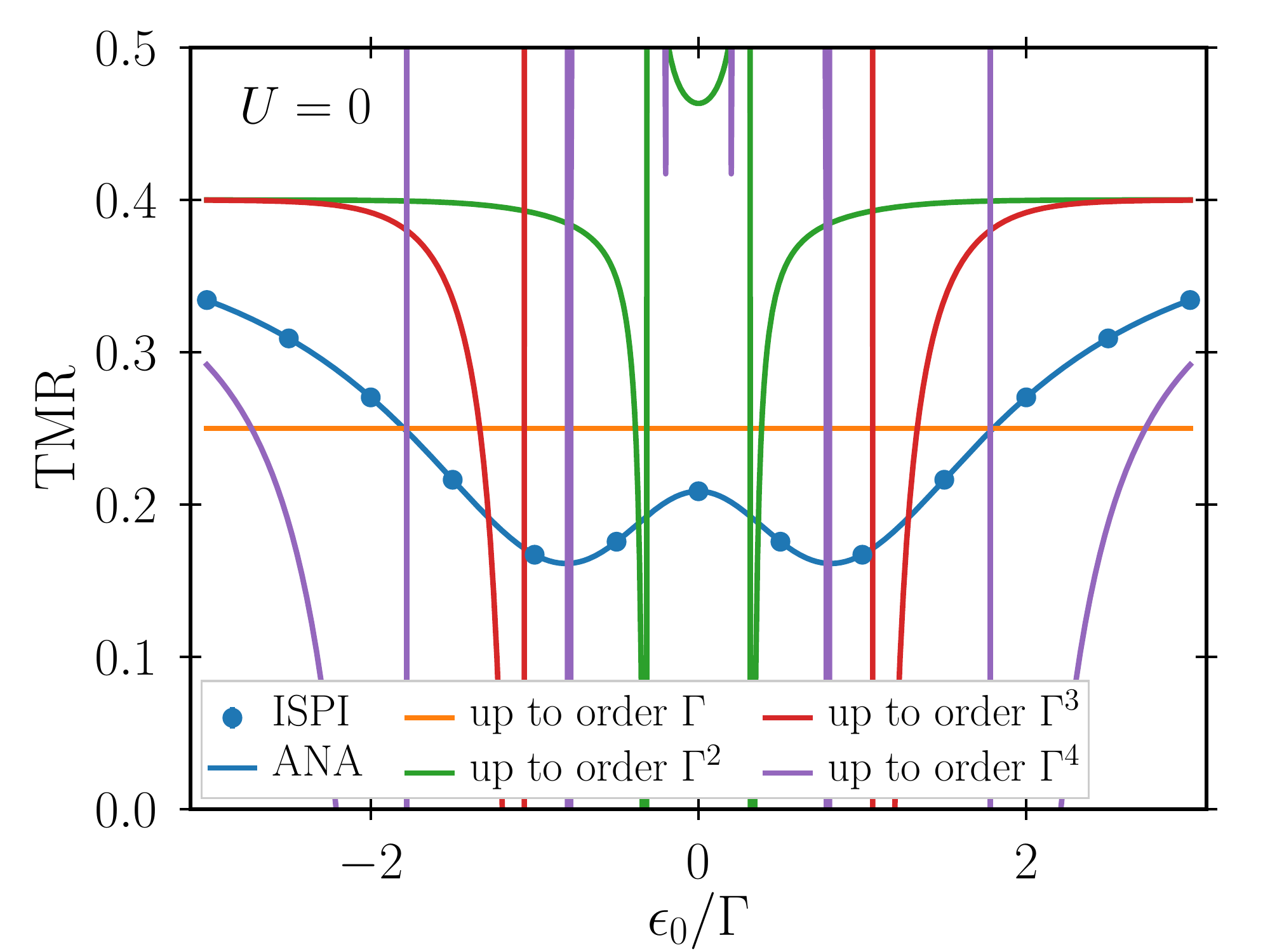}
\caption{\label{fig:highGam} Gate-voltage dependence of the TMR for the noninteracting case, $U=0$, at low temperature, $k_BT=0.2\Gamma$, in the linear-response regime, $eV=0.1\Gamma$. The leads' polarization was set to $p=0.5$. The ISPI data (blue dots) perfectly agree with the full analytical result (ANA). A perturbation expansion of the latter up to fourth order in $\Gamma$ serevely fails.}
\end{figure}

To demonstrate once more the failure of perturbation theory, we show in Fig.~\ref{fig:highGam} the gate-voltage dependence of the TMR for $U=0$ in the linear-response regime, $eV=0.1\Gamma$, and compare with the full analytical result (ANA) as well as a perturbative expansion of the latter up to fourth order in $\Gamma$.
The ISPI data (symbols) perfectly match with the full analytic result, with error bars smaller than symbol size, while finite-order perturbation theory is not only quantitatively but even qualitatively wrong.
This emphasizes again the importance of resonant-tunneling effects for the TMR.

\begin{figure}[ht!]
\centering
\includegraphics[width=\columnwidth]{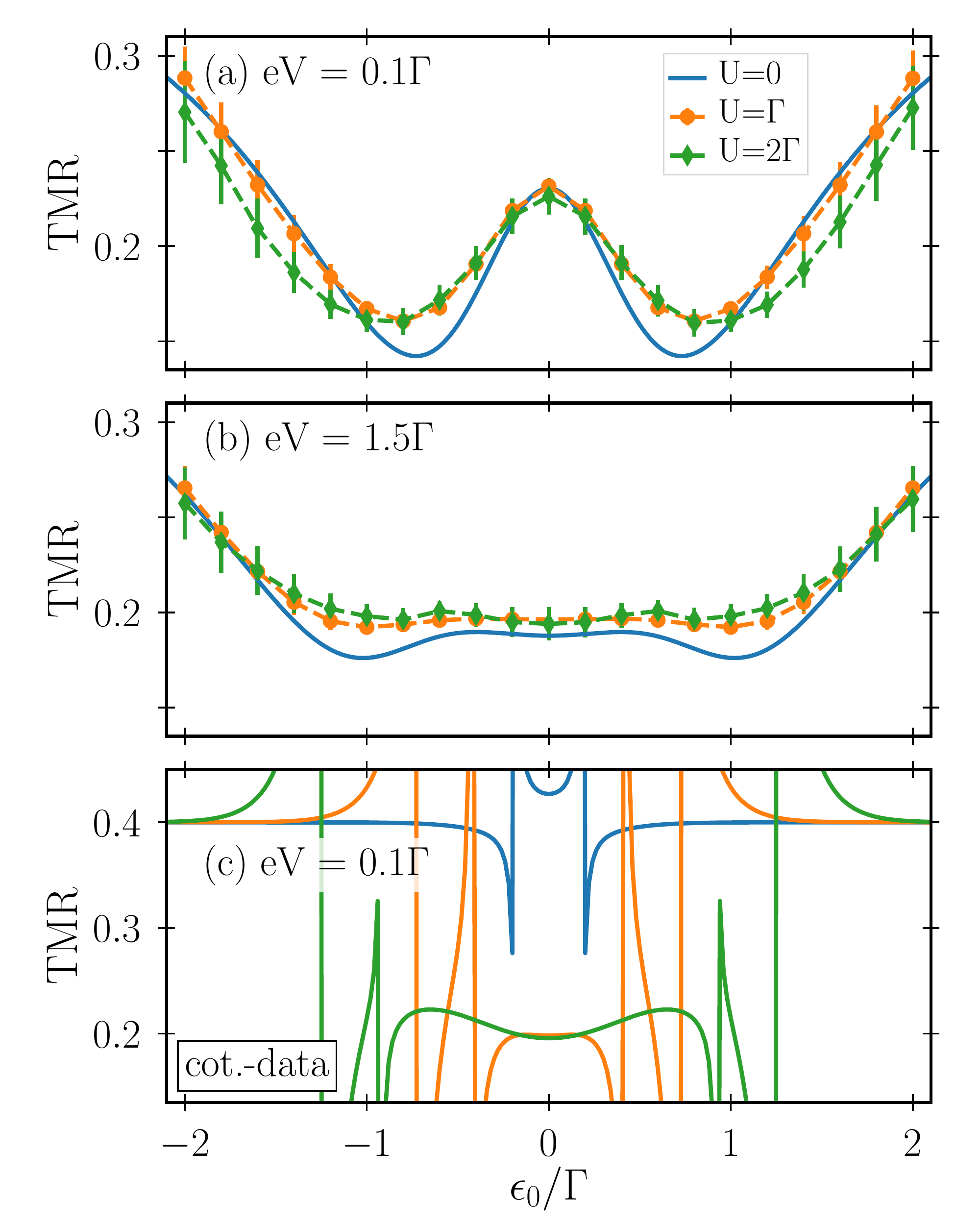}
\caption{\label{fig:TMRepsU} Gate-voltage dependence of the TMR for $U/\Gamma=0$, $1$, and $2$ and $k_BT=0.1\Gamma$. The polarization is $p=0.5$. The voltage is chosen to be (a) in the linear-response regime, $eV=0.1\Gamma$ and (b) slightly beyond the bifurcation point, $eV=1.5\Gamma$. In (c), we depict the results of a first- plus second-order calculation for the same parameters as in (a).}
\end{figure}

We now include finite Coulomb interaction $U \neq 0$.
In Fig.~\ref{fig:TMRepsU}(a), we show the gate-voltage dependence of the TMR for $U/\Gamma=0$, $1$, and $2$ in the linear-response regime, $eV=0.1\Gamma$. 
The local maximum at resonance, $\epsilon_0=0$, remains clearly visible for all chosen values of $U$.
The error bars of the ISPI data are quite moderate even for $U=2 \Gamma$.
Since for large $|\epsilon_0|$, the TMR goes up to Julliere's value, there are two minima symmetrically placed around the central maximum.
With increasing $U$, the depth of the minima seems to shrink a bit.

In panel (b) of Fig.~\ref{fig:TMRepsU}, we switch to the nonlinear-response regime, $eV=1.5\Gamma$. 
This bias voltage is chosen close to the bifurcation point indicated in Fig.~\ref{fig:TMRdens}.
The central maximum has just split into two local maxima, a finite Coulomb interaction $U$ seems to wash out the fine structure visible for $U=0$. 
Note that the limiting value of the TMR for $|\epsilon_0|\gg \Gamma$ is again set by Julliere's value. 

We complete the discussion of the gate-voltage dependence of the TMR with the remark that also for the case of finite $U$, a first- plus second-order calculation completely fails at low temperatures.
This is shown in panel (c) of Fig.~\ref{fig:TMRepsU}, in which the same parameters as in panel (a) are used.
 
\begin{figure}[ht!]
\centering
\includegraphics[width=\columnwidth]{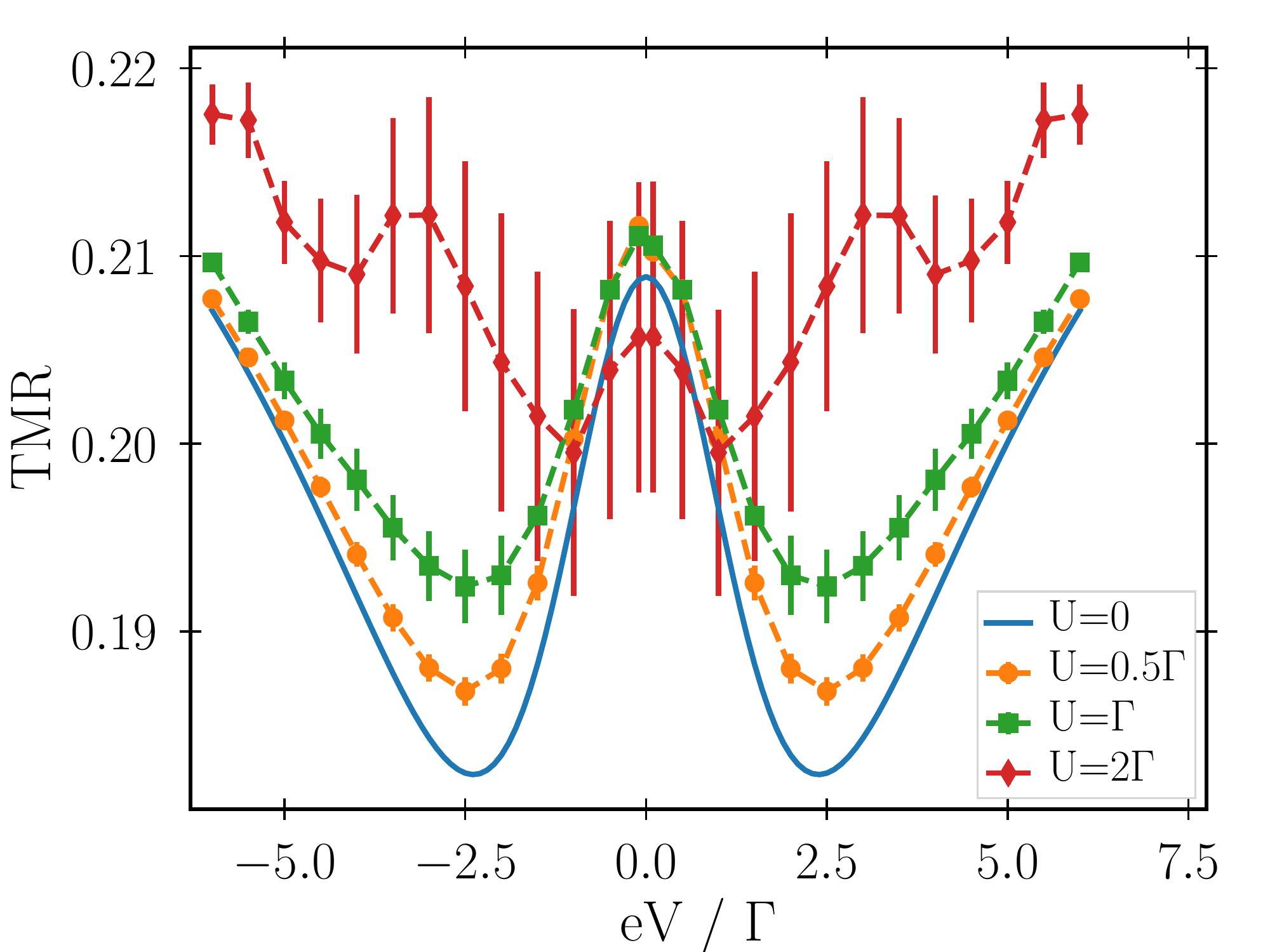}
\caption{\label{fig:TMRbias}Bias-voltage dependence of TMR for $U/\Gamma=0$, $0.5$, $1$ and $2$. 
The other parameters are $\epsilon_0=0$, $k_BT=0.2\Gamma$, and $p=0.5$.}
\end{figure}

Finally, we study the bias-voltage dependence of the TMR at $\epsilon_0=0$ and $k_BT=0.2\Gamma$ for different values of the Coulomb interaction strength, see Fig.~\ref{fig:TMRbias}.
Since the current is an antisymmetric function of $V$, the TMR is symmetric with respect to $V=0$. 
For large bias voltage, the TMR approaches the value obtained for sequential tunneling, $\text{TMR}=p^2=0.25$.

At $U=0$, there is a well-pronounced local maximum in the center.
Two local minima placed symmetrically around the central peak.
As $U$ increases, the form of the TMR becomes modified.
At the center (linear-response regime), the TMR first grows until $U\approx \Gamma$ and, then, drops again.
In the non-linear regime, $|eV|>\Gamma$, the interaction always increases the TMR.
As a consequence, the minima are getting washed out for $U> \Gamma$.
The error bars indicate the increasing numerical challenge for increasing $U$, especially in the linear-response regime.

\section{Conclusion and Outlook}
\label{sec:conclusion}

The tunnel magnetoresistance through a quantum-dot spin valve is a highly non-trivial function of temperature as well as gate and bias voltage.
For negligible Coulomb interaction, the TMR can be calculated analytically.
In the limit of weak tunnel coupling, a perturbative treatment of tunneling including sequential or sequential plus cotunneling is possible.
In a realistic experimental scenario, however, the energy scales for charging energy, tunnel strength, and temperature are all of the the same order of magnitude, and neither neglecting interactions nor a perturbative treatment of tunneling is justified.
In contrast, resonant-tunneling effects in the presence of finite Coulomb interaction are important.

To tackle the task of calculating the TMR for experimentally relevant transport regimes, we make use of an iterative summation of path integral (ISPI) scheme, which is based on the evaluation of nonequilibrium path integrals on the Keldysh contour in an iterative and deterministic manner. 
We have generalized this scheme to include spin-dependent tunneling.
ISPI naturally includes tunneling contributions from all orders in the tunnel-coupling strength, which seems to be extremely important for a proper determination of the TMR.
For intermediate values of the charging energy, up to the same order of magnitude as the tunnel-coupling strength, numerical convergence is achieved.

After benchmarking against known results for limiting cases, we have analyzed the temperature as well as the gate- and bias-voltage dependence of the TMR.
We find that for $k_B T \lesssim \Gamma$, perturbation theory severely fails since resonant tunneling becomes important.
The TMR displays a well-pronounced peak at $\epsilon_0=0$ and $V=0$. 
We are able to clearly resolve this peak within our ISPI scheme and to study the influence of a finite Coulomb interaction.

In conclusion, the ISPI treatment of the TMR through a quantum-dot spin valve offers a new theoretical tool that covers a transport regime in which other methods fail.
This includes the regime in which all the system parameters are of the same order and an expansion in a small parameter is impossible, which is particularly relevant in realistic experimental setups.

\section{Acknowledgements}
We acknowledge financial funding of the Deutsche Forschungsgemeinschaft (DFG, German Research Foundation) under project WE $5733/1-2$, KO $1987/5-2$ and – Projektnummer 278162697 – SFB 1242. Furthermore, we were inspired by fruitful discussions with Alfred Hucht on block matrix operations.

\appendix
\section{Current formulae for the noninteracting spin valve}
\label{appA}
In the absence of interaction, $U=0$, the current through a single-level quantum dot can be calculated analytically for all temperatures and bias voltages\cite{Jauho_1994}.
Including a finite spin polarization of the leads is straightforward.
The current formulae for a parallel ($p$) and antiparallel ($ap$) magnetization configuration of the leads are
\begin{widetext}
\begin{eqnarray}
I_p&=&\int\limits_{-\infty}^\infty d\omega \frac{2  \Gamma^2_\downarrow\Gamma^2_\uparrow+(\Gamma_\downarrow^2+\Gamma_\uparrow^2) (\epsilon_0 -\omega )^2}{\left(\Gamma_\downarrow^2+(\epsilon_0 -\omega )^2\right) \left(\Gamma_\uparrow^2+(\epsilon_0 -\omega )^2\right)}
\left[f\left(\omega -\frac{eV}{2}\right)-f\left(\omega+\frac{eV}{2}\right)\right]
\\
I_{ap}&=&\int\limits_{-\infty}^\infty d\omega\frac{8\Gamma_\downarrow\Gamma_\uparrow}{\left(\Gamma_\downarrow+\Gamma_\uparrow\right)^2+4(\epsilon_0 -\omega )^2}
\left[f\left(\omega -\frac{eV}{2}\right)-f\left(\omega + \frac{eV}{2}\right)\right] \, .
\end{eqnarray}
These integrals are evaluated with the help of Cauchy's residual theorem.
Collecting the residues at the poles of the Fermi functions (given by the Matsubara frequencies) and of the expression in front of the Fermi functions, we get
\begin{eqnarray}
I_p&=&\sum_{\sigma=\uparrow,\downarrow} \sum_{\gamma=\pm 1} \sum_{\gamma'=\pm 1}  \frac{\gamma}{2}\Gamma_\sigma \left[i \gamma' \Psi \left(\frac{1}{2}+\gamma'\frac{ \beta \Gamma_\sigma}{2 \pi} -\frac{i\beta}{2\pi}\left(\gamma\frac{e V}{2}-\epsilon_0\right)\right)-\pi f\left(\gamma\frac{e V}{2} + i\Gamma_{\sigma}+\epsilon_0\right)\right]
\\
I_{ap}&=&\sum_{\gamma=\pm 1} \sum_{\gamma'=\pm 1}  \gamma \Gamma (1-p^2)\left[i\gamma' \Psi \left(\frac{1}{2}+\gamma'\frac{\beta \Gamma}{2\pi}-\frac{i\beta}{2\pi}\left(\gamma\frac{e V}{2}-\epsilon_0\right)\right)-\pi f\left(\gamma \frac{e V}{2}+i\Gamma+\epsilon_0\right)\right] \, ,
\end{eqnarray}
where $\Psi(\ldots)$ denotes the digamma function.
\end{widetext}

\end{document}